# Minimal Interaction in the Local Landau Theory and Construction of the Phenomenological Ginzburg-Landau Potential in Terms of Electron-Phonon Interaction

A.Ya. Braginsky

Research Institute of Physics, Rostov State University, Rostov-on-Don

e-mail: **a.braginsky@mail.ru**

**Abstract**

In the present paper, the method for describing inhomogeneous states with local translational symmetry is proposed, based on the symmetry-dependent interaction between the order parameter (OP) and compensating field in the phenomenological Landau theory. It is shown that dimensionality of the compensating field in the extended derivative is associated with representation of the OP and does not coincide with the dimensionality of the derivative itself. This results in the correct definition of transformational properties of physical fields in the local Landau theory and in the equations of continuum theory of dislocations in the system of equations of state. The mechanism is proposed to construct the Ginzburg-Landau potential for states with local translational symmetry that correspond to HTS states. It is shown that minimal interaction between the superconducting OP and tensor compensating field is responsible for electron-phonon interaction in the BCS model.

PACS number(s):  64.60.Bb, **74.20.-z.**

**Introduction**

Gauge models with minimal interaction in the Landau theory were introduced in the pioneer paper by Ginzburg and Landau [1] and later in the De Gennes' model [2], in describing deformed SmA, and were borrowed from the field theory. In the present paper, we will show that, vice versa, minimal interaction is an integral part of the local Landau theory and is symmetry-dependent. Representations with $\vec{k} \neq 0$ are poorly described in the field theory, whereas subgroups of translations play significant role in Physics. Studies of nontrivial representations of the subgroup of translations provide the basis for crystallography and physics of phase transitions. Unlike field theory where compensating fields are determined by abstract gauge group and do not depend on representation of wave function with respect to the subgroup of translations (trivial representations of the translation subgroup with $\vec{k} = 0$ are usually considered in the field theory), here, in the present theory, transformational properties of compensating fields are entirely determined by local transformation properties of OPs. Hence we will show that, for the Ginzburg-Landau model, these local transformational properties are determined by the transformation of wave function under temporal translations:

$$\hat{\tau}\psi = e^{i\omega(x)\tau}\psi , \qquad (1)$$

while for the De Gennes' model, these are local transformational properties of the OP determined by spatial translations:

$$\hat{\vec{a}}\eta_l(\vec{X}) = e^{i\vec{k}_l(\vec{X})\vec{a}}\eta_l(\vec{X}) . \qquad (2)$$

An assumption concerning locality of transformational properties of the OP (dependence $\vec{k}_l = \vec{k}_l(\vec{X})$) was made in [3]; it is as allowable as the dependence of OP on the coordinates $\eta_l = \eta_l(\vec{X})$. This is true for the general case that in the inhomogeneous Landau theory, not only the value of OP may depend on coordinates, but the transformational properties of OP also depend on the coordinates (they are determined by the vector $\vec{k}$). Such OP describes an inhomogeneous low-symmetry state where in each macroscopically small region enumerated by the macrocoordinate $\vec{X}$ a vector $\vec{k}$ is defined, and in this point a local Landau potential $\Phi(\vec{X})$ can be constructed. Obviously dependence of the OP on $\vec{X}$ in the inhomogeneous models [1,2,4,5] implied that the local Landau potential existed. Otherwise, dependence of the OP on the coordinate would have been impossible since the OP is equivalent to the coefficients in the expansion of density of state into the Fourier series with respect to the coordinate.

In fact, the first attempt to examine the model with local transformational properties of OP in the Landau theory was made by de Gennes in [2]. Since the director in the deformed SmA (the single plane normal vector) depends on $\vec{X}: \vec{n} = \vec{n}(\vec{X})$, then the vector $\vec{k}$ of the smectic OP $\psi_{\vec{k}} = \psi_{\vec{k}}(\vec{X})$ also depends on $\vec{X}$, since $\vec{k} = \vec{n}/d$, where $d$ is the distance between the layers in SmA.

### 1. Compensating Field

For models with $\vec{k} \neq 0$ inhomogeneity $\vec{k}_l = \vec{k}_l(\vec{X})$ results in nontrivial transformations of the OP derivative under unit translations [3], [6].
In fact,

$$\hat{\vec{a}} \frac{\partial \eta_l}{\partial X_j} = \frac{\partial}{\partial X_j}\left(e^{i\vec{k}^l(\vec{x})\vec{a}} \eta_l(\vec{X})\right) = e^{i\vec{k}^l(\vec{x})\vec{a}} \left[ i \frac{\partial \vec{k}^l(\vec{X})\vec{a}}{\partial X_j} \eta_l(\vec{X}) + \frac{\partial \eta_l(\vec{X})}{\partial X_j} \right] \quad (3)$$

As it is known [4], the inhomogeneous Landau potential is an invariant function constructed on the OP's basis and its spatial derivatives with respect to symmetry of the high-symmetry phase. It follows from (3) that the translation operator transforms spatial derivatives of the OP, $\partial \eta_l / \partial X_j$, on the space of the OP itself, $\eta_l(\vec{X})$ (the first summand in the right side of expression (3)). Here, components of the OP are eigenfunctions with respect to the translation operators (2) because the subgroup of translations is an Abelian group, and it is known that irreducible representations (IR) of an Abelian group are one-dimensional. To construct invariants of the subgroup of translations which include spatial derivatives, we need first to construct the diagonal basis for the translation operator which includes spatial derivatives of the OP. Let us apply the procedure proposed in the gauge field theory and construct extended derivatives by introducing additional compensating fields in the derivative so that the extended derivatives would be eigenfunctions of the translation operator. Here, according to (3), compensating fields should be determined accurately within the gradient of the vector function. Let us show that this field is a tensor one and it is single.

Let us write down the extended derivative as:

$$D_j^l \eta_l = \left( \frac{\partial}{\partial X_j} - i\gamma \sum_p A_{pj}^l \right) \eta_l, \quad (4)$$

where the translation operator action on $\gamma A_{pj}^l$ is determined in such a way that the extended derivative (4) could be the eigenfunction of the translation operator:

$$\hat{\bar{a}}_q(\gamma A_{pj}^l) = \gamma A_{pj}^l + \delta_{pq}\frac{\partial k_p^l}{\partial X_j}a_q, \qquad (5)$$

Here, $A_{pj}^l$ is the compensating field, and $\gamma$ is a phenomenological charge. Dimensionality of the compensating field $A_{pj}^l$ is associated with dimensionality of the vector $\vec{k}^l$ and is a second-rank tensor (4), since $A_{pj}^l$ should undergo transformations like $\partial k_p^l/\partial X_j$ (5). Vectors in the star $\{\vec{k}\}$ are IR-dependent, since they are obtained from one vector $\vec{k}$ by operations from the point symmetry group. Hence it follows that one tensor field $A_{pj}$ can be chosen which will be compensating for all vectors $\vec{k}^l$ of the IR. For example, for the six-beam basis of the icosahedron [6]

$$\vec{k}_1 = (0,1,h), \qquad \vec{k}_2 = (0,\bar{1},h), \qquad \vec{k}_3 = (h,0,1), \vec{k}_4 = (h,0,\bar{1}), \qquad \vec{k}_5 = (1,h,0),$$

$$\vec{k}_6 = (\bar{1},h,0),$$

where $h = (\sqrt{5}+1)/2$

the extended derivatives take the form:

$$D_j^1\eta_1 = \left[\frac{\partial}{\partial x_j} - i\gamma(A_{2j} + hA_{3j})\right]\eta_1, \qquad D_j^2\eta_2 = \left[\frac{\partial}{\partial x_j} - i\gamma(-A_{2j} + hA_{3j})\right]\eta_2,$$

$$D_j^3\eta_3 = \left[\frac{\partial}{\partial x_j} - i\gamma(hA_{1j} + A_{3j})\right]\eta_3, \qquad D_j^4\eta_4 = \left[\frac{\partial}{\partial x_j} - i\gamma(hA_{1j} - A_{3j})\right]\eta_4, \quad (6)$$

$$D_j^5\eta_5 = \left[\frac{\partial}{\partial x_j} - i\gamma(A_{1j} + hA_{2j})\right]\eta_5, \qquad D_j^6\eta_6 = \left[\frac{\partial}{\partial x_j} - i\gamma(-A_{1j} + hA_{2j})\right]\eta_6.$$

$$\tilde{D}_j^l\tilde{\eta}_l = (D_j^l\eta_l)^*.$$

While constructing the compensating field in (6) we took into account that the change in the coordinates of the vector $\vec{k}(\vec{X})$ in reciprocal space $\mu_i(\vec{X})$ was equivalent to the change in the values of the basis vectors of the reciprocal space $\vec{k}(\vec{X}) = \mu_i\vec{b}_i(\vec{X})$. As a result, the extended derivatives $D_j^l\eta_l$ of the OP are eigenfunctions of the translation operator $\hat{\bar{a}}$ and are transformed, like the OP, into (2).

2. **Stress and Dislocations**

States described by OP with local translational properties $\vec{k} = \vec{k}(\vec{X})$, may be given illustrative interpretation: inhomogeneous deformation of crystal lattice, which is generally accompanied by distortions and occurrences of dislocations. Dislocations as linear incompatibilities of lattice occur at the boundaries of crystal regions which have different periods. Let us show that equations of state for the model with $\vec{k} = \vec{k}(\vec{X})$ contain equations of the elasticity theory for dislocations.

Using (4) we are able to construct a translation-invariant inhomogeneous Landau potential as a function of OP and its extended derivatives. In this model, the introduced tensor compensating field $A_{pj}$ is an independent variable, and variation of the potential with respect to it must be equal to zero. As in electrodynamics as well, we have to take into account the invariants of the tensor compensating field, which take the form of anti-symmetric derivatives with respect to the second index according to (5):

$$\sigma_{pj} = e_{jkn} \left( \partial A_{pn} / \partial X_k \right). \tag{7}$$

Physical interpretation for $A_{pj} \equiv \Sigma_{pj}$ is associated with the tensor potential of the stress field introduced by Kröner [7] who identified and described the analogy between magnetostatics and continuum theory of dislocations. The definition of the stress tensor (7) is the equilibrium condition for the solid state. Equations of state obtained from the variation of the local Landau potential with respect to the components of the compensating field $\delta \Phi / \delta A_{pj}$ coincide with the basic equations of the continuum theory of dislocations [8]

$$\rho_{pj} = e_{jkn} (\partial w_{pn} / \partial X_k). \tag{8}$$

Since $\partial \Phi / \partial \sigma_{pj} = w_{pj}$ is the tensor of elastic distortion and it follows herefrom, according to (8), that $\rho_{pj} = \partial \Phi / \partial A_{pj}$ is the density of dislocations, by definition [8]. In case of the icosahedral star, we will obtain, from (5), the expression for the density of dislocations:

$$\rho_{1j} = -i\gamma \left[ hH_j^3 - hH_j^{3*} + hH_j^4 - hH_j^{4*} + H_j^5 - H_j^{5*} - H_j^6 + H_j^{6*} \right],$$

$$\rho_{2j} = -i\gamma \left[ H_j^1 - H_j^{1*} - H_j^2 + H_j^{2*} + hH_j^5 - hH_j^{5*} + hH_j^6 - hH_j^{6*} \right], \tag{9}$$

$$\rho_{3j} = -i\gamma \left[ hH_j^1 - hH_j^{1*} + hH_j^2 - hH_j^{2*} + H_j^3 - H_j^{3*} - H_j^4 + H_j^{4*} \right],$$

where $H_j^l = \dfrac{\partial \Phi}{\partial(\partial \eta_l / \partial x_j)} \eta_l$, $H_j^{l*} = \dfrac{\partial \Phi}{\partial(\partial \eta_l^* / \partial x_j)} \eta_l^*$. $\tag{10}$

The expression (9) is a second-rank tensor at transformations from the point symmetry group of icosahedron. Indeed, transformations with respect to the second index are associated with three spatial derivatives. It is easy to verify that, the components $\rho_{pj}$ (9) are transformed with respect to the first index as a vector at transformations from the point symmetry group of icosahedron of the expressions (10). Thus, transformational properties of the OP determine tensor transformational properties of the observed values $\rho_{pj}$ (the expression (9) could as well be obtained as the first integral conjugate to the unit translations, by analogy with current density in electrodynamics).

Note that in no expression we specified explicit dependence of the potential on the invariants composed of the components of OP, compensating field and derivatives thereof. To obtain equations of state (8), it was sufficient to construct translational invariants and to require the potential to be a function of those invariants $\Phi(\vec{X}) = \Phi(\eta_l \eta_l^*, D_j^l \eta_l D_j^{l*} \eta_l^*, \sigma_{pi})$.

In the Abelian gauge model of the field theory, the change in the scalar phase of the wave function is compensated; the gauge transformation takes the form:

$$g\psi = e^{i\alpha(X)}\psi. \tag{11}$$

The compensating field is here transformed as a vector $g(eA_j) = eA_j + \partial\alpha/\partial X_j$. As opposed to the gauge model of the field theory (11), in the model (4) a tensor value (5) rather than a vector one was added to the extended derivative. Construction of the extended derivative in the Landau theory is implemented in such a way that the tensor in the extended derivative is transformed together with the OP at transformations from the point symmetry group (6). The compensating field enters the extended derivative not as a contraction of a tensor with respect to the first index, but as a tensor whose first index is transformed together with the vector $\vec{k}$ of the OP, while the second index is transformed together with the spatial derivatives. Thus in the model under consideration, the extended derivative is not a vector in terms of construction (4), (6). Note that we have arrived at this result by considering nontrivial representations of the subgroup of translations with $\vec{k} \neq 0$, in the inhomogeneous Landau theory, which are commonly not considered in the field theory due to finite dimensionality of IRs selected [9]. In the De Gennes model [2], in the general case, IRs were infinite-dimensional since at deviations of the director from the principal optic axis $\vec{n} = \vec{n}(\vec{X})$, the vector star $\{\vec{k}\}$ of the smectic OP is a surface of a cone. De Gennes made an attempt to construct a phenomenological potential for SmA similar to the Ginzburg-Landau potential in order to describe the effect of screening the stress field by elastic dislocations, similar to the Meissner effect [1]. However, he was selecting the compensating field to be a vector in the extended derivative based on the dimensionality of the derivative itself, and did not check the translational invariance of the constructed smectic potential. It is easily seen that the De Gennes' potential is not invariant with respect to the unit translation operator (3), since the vector field cannot compensate for the changes in the director $\vec{n}(\vec{X})$ in three dimensions. Vortex structure of tensor equations of state (7) results in the equations similar to London equations [1,10] and in the description of screening effect of the stress field by elastic dislocations. As it is known [2, 11], in the De Gennes model, that problem could not be solved because the Franck potential $\Phi = \Phi(\vec{n}(\vec{X}))$ for nematic was used therein to describe elastic properties of SmA which contained non-vortex summands of divergence-of-director type. The present model, containing tensor compensating field in the local Landau theory, is free of the above deficiencies [10].

## 3. Ginzburg-Landau Model

Let us consider, by analogy with spatial translations, the model with local transformational properties of the OP at temporal translations (1). In this case, the compensating vector field $A_j$ changes its sign at the inversion of time, by construction:

$$D_j\psi = \left(\frac{\partial}{\partial X_j} - ieA_j\right)\psi, \tag{12}$$

$$\hat{\tau}(eA_j) = eA_j + \frac{\partial\omega}{\partial X_j}\tau \tag{13}$$

(according to (13), the vector field $A_j$ is transformed as $\partial \omega / \partial X_j$).

In the Abelian field theory, change of the sign of the compensating field in the extended derivative is postulated [9] because, as mentioned above, a scalar phase in (11), which is invariant at temporary transformations, is the group gauge parameter for a model in electrodynamics. Knowing that the electromagnetic vector-potential $A_j$ changes its sign at the inversion of time, it is possible to state the contrary as well: such representation of the OP, for which the electromagnetic potential $A_j$ is acting as the compensating field, must have local transformational properties at temporary translations (1). Indeed, the extended derivative (12) contains the values that are transformed in different ways at the inversion of time $(\partial / \partial X_j - ieA_j)$, and the explanation must be given to this. Additional transformational properties of the compensating field in the extended derivative must be associated with the transformational properties of the OP. For example, the compensating field in the Ginzburg-Landau model is transformed together with the OP at the inversion of time $\hat{I}_\tau (\psi) = \psi^*$, being in agreement with the representation (1).

Thus, introduction of the local Abelian gauge group and additional determination of transformational properties of vector-potential at the inversion of time in the field theory [9] is equivalent to the Landau theory where OP has local transformational properties at temporary translations (1). Variation of the local Landau potential with respect to the components of the compensating field is in this case similar to variation of the Ginzburg-Landau potential with respect to the components of the electromagnetic potential and results in the classical Maxwell equations in the system of equations of state [1].

## 4. Superconductivity and Electron-Phonon Interaction

As it is known [12, 13], high-temperature superconductivity (HTS) states are inhomogeneous. However it may be assumed that in the HTS states short-range crystallographic order exists, and hence, a local Landau potential may be constructed. Hence, it is required to construct the Ginzburg-Landau potential for the states with $\vec{k} = \vec{k}(\vec{X})$. Representation of OP for the HTS state must have local transformational properties both (1) and (2). In this case, the extended derivative will include the linear summands of both tensor $A_{pj}$ and vector-potential $A_j$. Then, for the IR with $\vec{k} \neq 0$ and $\psi_{\vec{k}_l} = \psi_{\vec{k}_l(\vec{X})}$, let us assume that

$$\hat{\tau} \psi_{\vec{k}_l} = e^{i\omega(x)\tau} \psi_{\vec{k}_l}, \quad \hat{\tau} \psi^*_{\vec{k}_l} = e^{-i\omega(x)\tau} \psi^*_{\vec{k}_l}, \tag{14}$$

and with inversion of time $\hat{I}_\tau (\psi_{\vec{k}_l}) = \psi_{\vec{k}_l}^*$ for each $\vec{k}_l$. Taking nontrivial transformations of OP into account at temporary translations will result in increase in the IR dimensionality two times $\{\psi_{\vec{k}_l}, \psi^*_{\vec{k}_l}\}$. In this case, quadratic invariants of the OP take the form:

$$\sum_l \psi_{\vec{k}_l} \psi^*_{-\vec{k}_l} + \psi^*_{\vec{k}_l} \psi_{-\vec{k}_l}. \tag{15}$$

Extended derivatives, according to (4), (12):

$$D_j \psi_{\vec{k}_l} = \left( \frac{\partial}{\partial X_j} - i\gamma \sum_p A^l_{pj} - ieA_j \right) \psi_{\vec{k}_l}, \quad D_j \psi^*_{\vec{k}_l} = \left( \frac{\partial}{\partial X_j} - i\gamma \sum_p A^l_{pj} + ieA_j \right) \psi^*_{\vec{k}_l}, \tag{16}$$

$$D_j\psi_{-\vec{k}_l} = \left(\frac{\partial}{\partial X_j} + i\gamma\sum_p A^l_{pj} - ieA_j\right)\psi_{-\vec{k}_l}, D_j\psi^*_{-\vec{k}_l} = \left(\frac{\partial}{\partial X_j} + i\gamma\sum_p A^l_{pj} + ieA_j\right)\psi^*_{-\vec{k}_l}$$

and hence, quadratic gradient invariants from (15), (16) which are responsible for minimal interaction, may be written in the form:

$$\sum_{jl} D_j\psi_{\vec{k}_l} D_j\psi^*_{-\vec{k}_l} + D_j\psi^*_{\vec{k}_l} D_j\psi_{-\vec{k}_l}. \tag{17}$$

In fact, the extended derivative (16) contains the components of 4-tensor, the first three of them compensating for inhomogeneous changes in the OP at spatial translations, while the fourth one doing this at temporary translations. It is obvious from (16) and (17) that $\partial\Phi/\partial A_j$ have dimensionality of the current vector. London equations in the model (14) also stem from (16), (17). Indeed, the current vector only contains members, linear with respect to $A_j$, with coefficients $\psi_{\vec{k}_l}\psi^*_{-\vec{k}_l} + \psi^*_{\vec{k}_l}\psi_{-\vec{k}_l} = 2|\psi|^2$, and does not contain $A_{pj}$, since the components of $\sum_p A^l_{pj}$ enter into the London equations with zero coefficients: $\psi_{\vec{k}_l}\psi^*_{-\vec{k}_l} - \psi^*_{\vec{k}_l}\psi_{-\vec{k}_l} = 0$.

Hence, in the present model (14), we observe doubling of the coefficient before the electromagnetic potential in the London equation, that is corresponding to the idea of electron pairing in the BCS model; the vector of current density does not depend on the potential of the stress field, and this very fact means the absence of dissipation in the superconducting state. Since, in the absence of OP, the equations of state $\delta\Phi/\delta A_{pj} = 0$ correspond to the equations of the theory of elasticity [6], and dynamic equations for the free field $A_{pj}$ are equivalent to wave equations for the electromagnetic potential, then it is clear that the tensor $A_{pj}$ corresponds to the phonon potential in the BCS model and is responsible for electron-phonon interaction (16) in the superconducting state. Here, using the term "phonon potential", we imply that the tensor $A_{pj}$ describes elastic properties of the crystal lattice. Its conjugate value, $\partial\Phi/\partial A_{pj} = \rho_{pj}$, tensor of density of dislocations, initiates, in the inhomogeneous state, internal stress. Here we identify the perfect analogy to the current and electromagnetic field which is produced thereby.

In this model, there is a duality in the choice of definitions of physical fields [14]. It consists in the fact that if one of the quantities (7), (8) defined as an invariant subgroup of translations, the other is determined from the equations of state. Compensating field can be defined as the tensor of elastic distortion $A_{pj} \equiv w_{pj}$. Then $\partial\Phi/\partial A_{pj} = \sigma_{pj}$. This means that the internal stresses are a source of phonons, and the dislocation density is defined in a continuous medium and is independent of the strain tensor. Note that the choice of definition of vortex invariants do not affect the main conclusions of this work.

. Thus, the phenomenological model (16) agrees with the BCS model and describes not only electromagnetic interaction but electron-phonon interaction as well. The latter being taken into account results in electron pairing, and in the superconducting state where the current does not depend on the internal stress. Internal stress in the present model does not depend on current, but rather is determined by external conditions.

Extension of the derivative associated with the introduction of the phonon tensor potential $A_{pj}$ in (16) is symmetry-dependent on the local translational symmetry in the HTS specimens. In states with local translational symmetry, inhomogeneous distribution of electron density results in occurrence of internal stress which is the source of phonons. It should be expected that in such states electron pairing will take place at higher temperature, this actually taking place in HTS states.

For ordinary superconductors, electron-phonon interaction should also be present in phenomenological descriptions of inhomogeneous states. It takes into account interaction between current and lattice deformation, but inhomogeneity of state is, in this case, the result of inhomogeneous distribution of the superconducting OP in the external magnetic field.

Let us focus on the fact that in order to double the coefficient before the electromagnetic potential in the London equations associated with electron pairing, we did not need to double the charge in the extended derivative and to re-normalize the wave function, as it was done in [15]. Appropriate selection of representation with $\vec{k} \neq 0$ (14) allows settling this problem in the phenomenological theory.

Therefore, it follows from the requirement of translational invariance of the local Landau potential that, in describing superconducting states, the Ginzburg-Landau potential must compulsorily take into account electron-phonon interaction, this being in agreement with the BCS theory.

**Conclusions**

In the proposed model of the Landau theory, the symmetry group is global (an ordinary subgroup of translations of space-time, probably even a discrete one), whereas the representation is local. In the gauge field theory, the abstract gauge symmetry group itself is local. The notion of local gauge group was introduced to provide substantiation for minimal interaction in electrodynamics. As can be seen from the above considerations, the concept of locality of the transformational properties of OP with respect to the global subgroup of translations of space-time is a better tool to handle this problem since it entirely defines transformational properties of physical fields and does not require abstract symmetry groups to be introduced. Non-Abelian Young-Mills [16] fields are not suitable as compensating fields in the Landau theory, because they assume their values in the abstract internal space which is not linked to space-time, by definition. Young-Mills fields are not second-rank tensors; hence the equations (7), (8) cannot be obtained in a non-Abelian gauge theory. In the proposed model, tensor compensating fields do not act in the space of OP functions, as Young-Mills fields do, rather, they are transformed together with the OP, or more exactly, with its vector $\vec{k}$, and their dimensionality is not associated with dimensionality of the IR.

**Acknowledgements**

The author wishes to acknowledge the supervisor of his studies, Yu.M.Gufan, for invaluable discussions and support without which the present paper would be impossible.

**References**

1. V.L. Ginzburg and L.D.Landau, Zh. Exp. Teor. Fiz. **20**, 1064 (1950).
2. P.G.De Gennes Solid State Commun. **10**, 753-756 (1972).
3. A.Ya. Braginsky, Fiz. Tverd. Tela (St. Petrsburg) 32, 10 (1990) [Sov. Phys. Solid State 32, 4 (1990)]
4. E.M. Lifshitz, Zh. Eksp. Teor. Fiz. **11**, 255 (1941); Zh. Eksp. Teor. Fiz. **11**, 269 (1941).
5. I.E. Dzyaloshinskii, Zh. Eksp. Teor. Fiz. **46,** 1420 (1964)[Sov. Phys. JETP **19**, 960 (1964)]
6. A.Ya. Braginsky, Zh. Eksp. Teor. Fiz. **132,** 30 (2007) [JETPH **105,** 40 (2007)].
7. R. De Wit   Solid State Physics **10**, New York, 249 (1960).
8. L.D.Landau and E.M.Lifshits, Course of Theoretical Physics, Vol.7: Theory of Elasticity 4$^{th}$ ed. (Nauka, Moscow, 1987; Pergamon, New York, 1986).
9. N.N. Bogolubov and D.V. Shirkov, An Introduction to the Theory of Quantized Fields (Wiley-Interscience, New York 1980).
10. A.Ya Braginsky Physical Revew **B 67**, 174113 (2003).
11. B.I. Halperin , T.C. Lubensky   Solid State Commun. **14**, 997-1001 (1974).
12. S.H. Pan et al.   Nature **413**, 282-285 (2001).
13. D.Larbalestier, A.Gurevich, D.M. Feldman, A. Polyanskii   Nature **414,** 368-377 (2001).
14. A.Ya Braginsky   Physical Revew B 66, 054202 (2002).
15. E.M. Lifshitz and L.P. Pitaevsky Course of Theoretical Physics, Vol.9: Statistical physics: pt. 2: Theory of the condensed state (Pergamon, New York, 1980).
16. L.D. Fadeev and A.A. Slavnov, Gauge Fields: Introduction to Quantum Theory (Nauka, Moscow 1978; Addison-Wesley, Redwood City, CA, 1990).